\documentclass[aip,jcp,reprint,showpacs,amsmath,amssymb]{revtex4-1}
\usepackage{graphicx}

\begin{document}

\title{Structure and dynamics in yttrium-based molten rare earth alkali fluorides}

\author{Maximilien Levesque}
\email{maximilien.levesque@gmail.com}
\affiliation{Department of Materials, University of Oxford, Parks Road, Oxford
OX1 3PH, United Kingdom}

\author{Vincent Sarou-Kanian}
\affiliation{CNRS, UPR3079, CEMHTI, 1D avenue de la Recherche Scientifique, 45071
Orl\'eans cedex 2, France}
\affiliation{Facult\'e des sciences, Universit\'e d\textquoteright{}Orl\'eans, avenue
du Parc Floral, BP 6749, 45067 Orl\'eans cedex 2, France}

\author{Mathieu Salanne}
\email{mathieu.salanne@upmc.fr}
\affiliation{UPMC Univ Paris 06, CNRS, ESPCI, UMR 7195, PECSA, F-75005, Paris, France}

\author{Mallory Gobet}
\affiliation{CNRS, UPR3079, CEMHTI, 1D avenue de la Recherche Scientifique, 45071
Orl\'eans cedex 2, France}
\affiliation{Facult\'e des sciences, Universit\'e d\textquoteright{}Orl\'eans, avenue
du Parc Floral, BP 6749, 45067 Orl\'eans cedex 2, France}
\affiliation{Hunter College of the City University of New York, Department of Physics \& Astronomy, New York, NY 10065, USA}

\author{Henri Groult}
\affiliation{UPMC Univ Paris 06, CNRS, ESPCI, UMR 7195, PECSA, F-75005, Paris, France}

\author{Catherine Bessada}
\affiliation{CNRS, UPR3079, CEMHTI, 1D avenue de la Recherche Scientifique, 45071
Orl\'eans cedex 2, France}
\affiliation{Facult\'e des sciences, Universit\'e d\textquoteright{}Orl\'eans, avenue
du Parc Floral, BP 6749, 45067 Orl\'eans cedex 2, France}

\author{Paul A. Madden}
\affiliation{Department of Materials, University of Oxford, Parks Road, Oxford
OX1 3PH, United Kingdom}

\author{Anne-Laure Rollet}
\affiliation{UPMC Univ Paris 06, CNRS, ESPCI, UMR 7195, PECSA, F-75005, Paris, France}

\begin{abstract}
The transport properties of molten LiF-YF$_3$ mixtures have been studied by pulsed field gradient nuclear
magnetic resonance spectroscopy, potentiometric experiments, and molecular dynamics simulations. The calculated diffusion coefficients and electric conductivities compare very well with the measurements across a wide composition range. We then extract static (radial distribution functions, coordination numbers distributions) and dynamic (cage correlation functions) quantities from the simulations. Then, we discuss the interplay between the microscopic structure of the molten salts and their dynamic properties. It is often considered that variations in the diffusion coefficient of the anions are mainly driven by the evolution of its coordination with the metallic ion (Y$^{3+}$ here). We compare this system with fluorozirconate melts and demonstrate that the coordination number is a poor indicator of the evolution of the diffusion coefficient. Instead, we propose to use the ionic bonds lifetime. We show that the weak Y-F ionic bonds in LiF-YF$_3$ do not induce the expected tendency of the fluoride diffusion coefficient to converge toward the one of yttrium cation when the content in YF$_3$ increases. Implications on the validity of the Nernst-Einstein relation for estimating the electrical conductivity are discussed.
\end{abstract}
\maketitle

\section{Introduction}

Molten salts are liquid mixtures of ionic species at high temperatures. They receive much attention due to their diverse but promising applications
that range from metal production~\cite{groult2008a,groult2011a,cassayre2010a}, the pyrochemical treatment of nuclear waste~\cite{hamel2007a,salanne2008a,taxil2009a,fukasawa2012a}, primary or secondary coolants in several concepts of generation IV fission~\cite{waldrop2012a} and fusion~\cite{ihli2008a} nuclear reactors, and electricity storage devices~\cite{bradwell2012a}. 

In the context of nuclear applications, the nuclear reactions produce rare earth elements in the core. From the neutronic point of view, these have a poisoning effect and must therefore be eliminated.
One of the advantages of most of the concepts of reactors based on molten salts, including the promising thorium-fueled molten salt fast reactor (MSFR) is that their fuel is dissolved in the primary coolant (which is the molten salt)~\cite{mathieu2006a,delpech2009a}.
Such liquid state fuel allows for the elimination and reprocessing of several fission products and rare earth elements during the reactor operation, via chemical or electrochemical methods~\cite{gibilaro2009a,chamelot2010a}. One drawback is the complex handling of the molten salt fuel inside the (recycling) circuit: important care must be taken for controlling the physico-chemical conditions\cite{levesque2012b,levesque2012c}, and additional efforts have to be made on the development of adapted structure materials~\cite{waldrop2012a,levesque2012a,levesque2013a}.

The knowledge on molten fluoride salts chemistry has long remained fragmented and limited to a few experimental techniques (mainly electrochemical ones) due to experimental hindrances. One has to deal with high temperatures (from 500~K to 1800~K) and to control the corrosive properties by setting up original experiments. Many efforts have been made towards this direction and several experimental techniques have been adapted to the study of molten salts. The development of new cells has unlocked the study of molten fluorides by EXAFS spectroscopy~\cite{rollet2004a}, which, coupled to other spectroscopic methods such as nuclear magnetic resonnance (NMR)~\cite{lacassagne2002a,bessada2006a} or Raman~\cite{toth1973a,dracopoulos2000a}, is able to provide a view of the local structure in these melts. As for the dynamic properties, similar improvements have been made and the self-diffusion coefficients can be measured up to 1500~K by pulsed field gradient nuclear
magnetic resonance (PFG-NMR)~\cite{rollet2009a,rollet2010a}. Such measurements are important since the diffusion coefficients provide an information on the {\it individual} dynamics of the ions, complementing the viscosity or electrical conductivity measurements which inform about the {\it collective} dynamics of the macroscopic system. 

During the past decade, these experimental efforts have been accompanied by the development of simulation tools adapted to the study of molten fluorides and chlorides, namely density functional theory~\cite{benes2009a}, thermodynamic modeling~\cite{benes2009b,robelin2012a,dewan2013a} and molecular dynamics (MD)~\cite{salanne2007b,ohtori2009a}. Among those, the latter has proven to be efficient for the study of the structure and dynamics of the system at the microscopic scale, allowing the interpretation of experimental results~\cite{salanne2006a,saroukanian2009a,pauvert2011a} and the prediction of quantities that hitherto remained unknown~\cite{dewan2013a}. The simulations are based a polarizable ion model, which has strong physical grounds~\cite{salanne2011c,madden1996a,madden2006a} and can efficiently be parameterized from first-principles~\cite{salanne2012b,aguado2003b,rotenberg2010a}, \textit{i.e.}, without having empirically recourse to any experimental data.

In this work, we combine state-of-the-art experimental and simulation tools to
study a series of LiF-YF$_{3}$ molten salts of varying compositions. We first describe shortly the methodological aspects. Then, we compare measured and calculated self-diffusion
coefficients and electrical conductivity to show that MD simulations are in
quantitative agreement with experiments, which allows for further use of the simulations
to discuss the interplay between the microscopic structure of the electrolytes and their dynamic properties.

\section{Methods}

\subsection{Pulsed field gradient nuclear magnetic resonance}

Mixtures of LiF and YF$_3$ salts (purity $99.99$~\%)
were prepared in a glove box under argon in order to avoid H$_{2}$O and O$_{2}$ contamination
of the samples. We use boron nitride (BN) crucibles to confine the salts, and to prevent them from any chemical reaction.
The amount of salt in each crucible is
\textit{ca}.~50~mg. The BN crucible is heated by a symmetrical irradiation
with two CO$_{2}$ lasers. The laser power is increased progressively until the target temperature is reached.

The high temperature PFG NMR spectra were recorded using a Bruker Avance
WB 400~MHz operating at 9.40~T. An argon stream prevents oxidation of the
BN at high temperature. The 10~mm liquid NMR probe is specially designed by Bruker and adapted in CEMHTI to work up
to 1500~K. It is equipped with a gradient coil providing 5.5~G$\cdot$cm$^{-1}\cdot$A$^{-1}$
that is combined with a gradient amplifier of 10~A. We use a NMR pulse sequence combining bipolar gradient pulses and stimulated
echo\cite{Cotts1969}. It is repeated with 8 gradients of linearly increasing strength.

Measurements were performed using the following NMR parameters: durations
of radiofrequency magnetic field application for $\pi/2$ pulses $p90=13$~$\text{\ensuremath{\mu}}$s
($^{19}$F) and $p90=9$~$\text{\ensuremath{\mu}}$s ($^{7}$Li),
gradient strength $g$ was varied from 0 to 50~G/cm, gradient application
time $\delta$ between 1 and 5~ms. For $^{7}$Li measurements, we
used a pre-saturation cycle before the diffusion sequence because
of the long relaxation time $T_{1}$ of this nucleus (25~s
in pure molten LiF).

Finally, the self-diffusion coefficients were obtained by nonlinear least-square
fitting of the echo attenuation\cite{Cotts1969}.

\subsection{Electrical conductivity potentiometric measurements}

The experimental setup used for measuring the electrical conductivity is described in details in reference~\cite{rollet2012b}; the conductivity cell configuration was derived from the one proposed by Hives and Thonstad~\cite{hives2004a}. The conductivity was measured for LiF-YF$_3$ mixtures with compositions ranging from $x_{\rm YF_3}$~=~0.00 to 0.18, at several temperatures. Here we report the results obtained for the 1130~K isotherm. All the details are given in the Supplementary Material Document No.1~\cite{si}.

\subsection{Molecular dynamics simulations}

We use molecular dynamics (MD) to compute static and dynamic properties of the molten
salt at the atomic scale. The MD simulations used therein rely on interatomic interaction potentials parameterized from first principles. The parameterization procedure is described elsewhere~\cite{salanne2012b}. They are of \textit{ab initio}
accuracy and have been successful
in reproducing accurately the electrical conductivity, the heat capacity,
the viscosity, etc, of various ionic materials (mainly molten fluorides and solid oxides\cite{burbano2012a,marrocchelli2012a}). In the case
of molten fluorides, the interaction potential is best described as the sum of
four different components: charge-charge, dispersion, overlap repulsion,
and polarization. The detailed description of the potential is given in the Supplementary Material Document No.1~\cite{si}.

We studied 11 concentrations ranging from $x_{\rm YF_3}$~=~0.10 to 0.60
by steps of 0.05. Each system was simulated along two isotherms. The first one at 1130~K allows for comparison with experimental data. The other one at 1200~K allows for comparison with previous works on fluorozirconates~\cite{pauvert2010a,pauvert2011a}. For both temperatures, (i) the equilibrium volume was determined from a simulation in the isothermal–isobaric (NPT) ensemble with a target pressure of 1 atm, using the method described by Martyna \textit{et al.}~\cite{martyna1994a}; (ii) production runs were then performed in the canonical (NVT) ensemble,~\cite{martyna1992a} with a time step of 1.0~fs. The total simulation time was of 4~ns for the production runs for each system. The relaxation time of the thermostat was set to 20~ps. The simulation cell parameters and the number of atom of each species are summarized for each composition and temperature in the Supplementary Material Document No.1~\cite{si}. The cut-off distance for the real space part of the Ewald sum~\cite{aguado2003a} and the short range potential was set to half the length ($L$) of the cubic simulation cell. The value of the convergence parameter in the Ewald sum was set equal to $5.6/L$ and 8$^3$ $k$-vectors were used for the reciprocal space part calculation of the forces and potential energy.

\section{Transport properties}

\begin{figure}
\includegraphics[width=1\columnwidth]{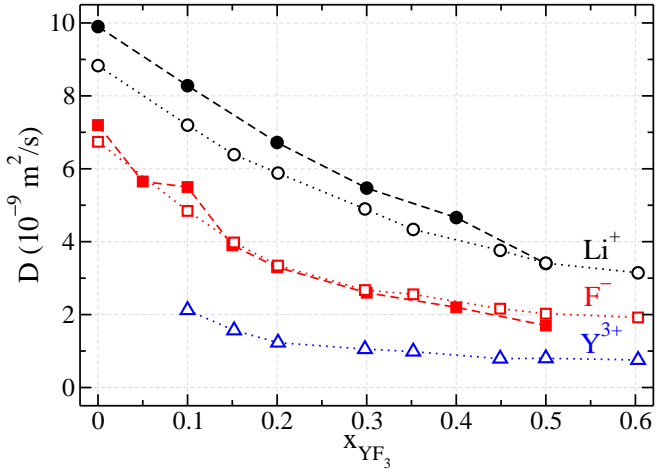}

\caption{Self-diffusion coefficients of lithium (black circles), fluoride (red squares) and yttrium (blue triangles) ions in LiF-YF$_3$ at $1130\pm10$~K
as a function of the mole fraction in YF$_3$. Full (open) symbols indicate experimental (Molecular Dynamics) data\label{fig:diffusion_coef}.
Dashed and dotted lines are just guides for the eyes.}
\end{figure}

The self-diffusion coefficients of fluoride, lithium and
yttrium ions along the 1130~K isotherm are presented
in Fig.~\ref{fig:diffusion_coef} as a function of the salt concentration.
Both experimental and molecular dynamics results are presented
for F$^-$ and Li$^+$. For Y$^{3+}$, the self-diffusion coefficient can only be measured from molecular simulations, due to the NMR characteristics of $^{89}$Y (low gyromagnetic ratio and long relaxation time up to 45~min). The self-diffusion coefficient $D_{\alpha}$ of species $\alpha$
is extracted from molecular dynamics simulations using Einstein's relation,
\textit{i.e.}, from the long-time slope of mean squared displacement:
\begin{equation}
D\left(\alpha\right)=\frac{1}{6}\lim_{t\rightarrow\infty}\partial_t  \left\langle\left|r_{i}(t)-r_{i}\left(0\right)\right|^{2}\right\rangle,
\end{equation}
where $r_{i}\left(t\right)$ is the position of atom $i$ of type $\alpha$
at time $t$. The brackets denote an ensemble average over all trajectories and atoms $i$ of type $\alpha$. The self-diffusion of Li$^+$ is about 60~\% higher than
that of F$^-$, whatever the concentration. Excellent agreement is found between
experiments and simulations. We can therefore be confident in the
MD simulation results for Y$^{3+}$. The latter value is about one third of the diffusion
coefficient of Li$^+$ over the whole range of compositions. All self-diffusion coefficients decrease at the same rate with the increase in yttrium fluoride concentration.

\begin{figure}
\includegraphics[width=1\columnwidth]{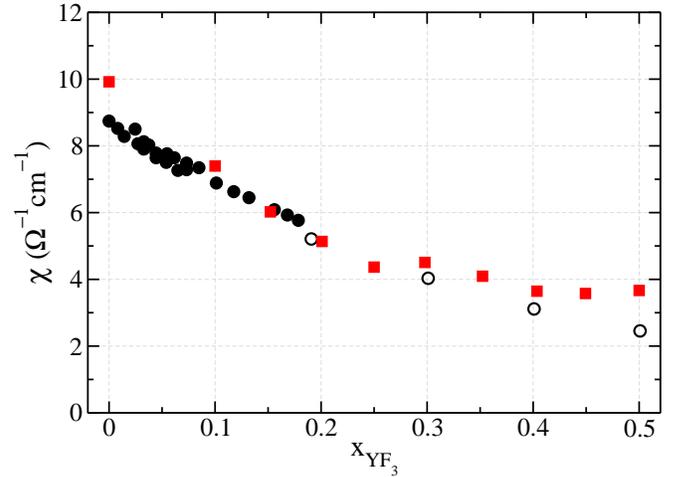}
\caption{Electrical conductivity $\chi$ of molten LiF-YF$_{3}$ mixtures as a function of the
mole fraction in YF$_3$ at 1130~K. (black circles) experimental data from
this study; (white circles) experimental data from reference~\cite{janz1988a}; (red squares)
molecular dynamics results.\label{fig:conductivity}}
\end{figure}

The electrical conductivities measured
at a temperature of 1130~K are presented in
Fig.~\ref{fig:conductivity} for different salt concentrations in YF$_3$ and compared to experimental results\cite{janz1988a}.
In molecular dynamics simulations, the electrical conductivity is calculated from
\begin{equation}
\chi=\frac{\beta e^2}{6V}\lim_{t\rightarrow\infty} \partial_t \left\langle \left|\sum_{\alpha}^{N_{\alpha}}q_{\alpha}\Delta_{\alpha}\left(t\right)\right|^{2}\right\rangle ,\label{eq:conductivity}
\end{equation}
where $\beta= 1/k_BT$ with $k_B$ the Boltzmann constant, $e$ is the elementary charge,
 $V=L^3$ the volume of the cubic simulation cell of length $L$, and
$\Delta_{\alpha}\left(t\right)=\sum_{i\in\alpha}\delta r_{i}\left(t\right)$
is the net displacement of all the ions of species $\alpha$ and charge
$q_{\alpha}$ in time $t$. There are $N_{\alpha}=3$ species here.
The error bars associated to our calculations of the electrical conductivity
from Eq.~\ref{eq:conductivity} are estimated to 10~\%. Such high values are
 due to the collective nature of $\chi$. It is not
averaged over both the time steps and all the atoms of a given species, as self-diffusion coefficients are, and consequently require long simulations runs. The true electrical conductivity of Eq.~\ref{eq:conductivity} is thus often approximated
by the Nernst-Einstein formulae, which relates $\chi$ to the individual
diffusion coefficients of the various species in the liquid. Surprisingly, we find the electrical
conductivity calculated from the Nernst-Einstein formulae to be overestimated by only 
10 to 20~\%. Larger differences, typically 40~\%, are usually observed for such salts\cite{borucka_self-diffusion_1957,harris_relations_2010}. Margulis et al. showed that the Nernst-Einstein relation always  overestimates the conductivity in molten salts and ionic liquids because of conditions in the momentum balance that contraint the motion of ions of same charge to be anticorrelated\cite{kashyap_how_2011}. The real electrical conductivity, as calculated from MD using Eq.~\ref{eq:conductivity}, is plotted
in Fig.~\ref{fig:conductivity}. Again, good agreement is found with experiments.

The electrical conductivity falls from $\approx9$~$\Omega^{-1}$cm$^{-1}$
in pure LiF to $\approx3$~$\Omega^{-1}$cm$^{-1}$,when the
molar fraction of YF$_{3}$ rises from 0 to 0.5. This decrease in
$\chi$ despite the addition of the highly charged Y$^{3+}$ ions is not surprising in view of the variations observed in other mixtures~\cite{salanne2006a,salanne2007b,janz1988a}.
It correlates with the reduced diffusion of all
species with $x_{\text{YF}_{3}}$ observed above; in the following
sections, we will analyze these variations in light of the microscopic structure of the melt.

\section{Microscopic structure}

In the previous section, we have shown that our molecular dynamics simulations were able to predict the transport properties of molten LiF-YF$_3$ with a very high accuracy. An advantage of such simulations is that we can extract precisely atomic scale informations on the
local structure of such melts. We will mainly focus on the study of the solvation shell of yttrium ions.

\begin{figure}
\includegraphics[width=1\columnwidth]{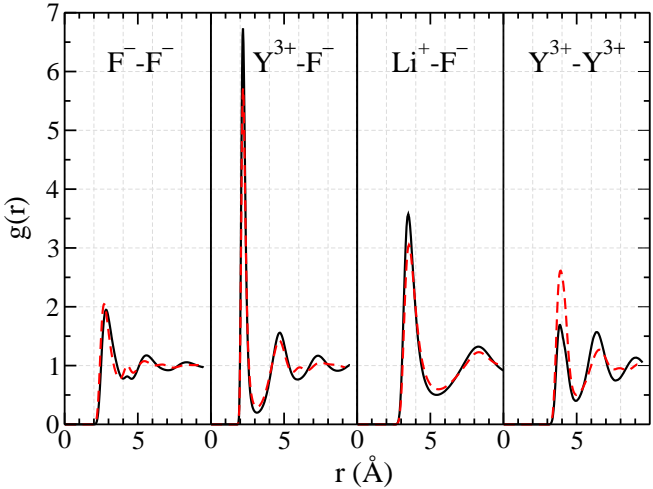}
\caption{\label{fig:rdf}
Radial distribution functions in LiF-YF$_3$ at 1130~K for $x_{\text{YF}_{3}}=0.15$
(black line) and 0.60 (red dashed line).}
\end{figure}

\begin{figure}
\includegraphics[width=1\columnwidth]{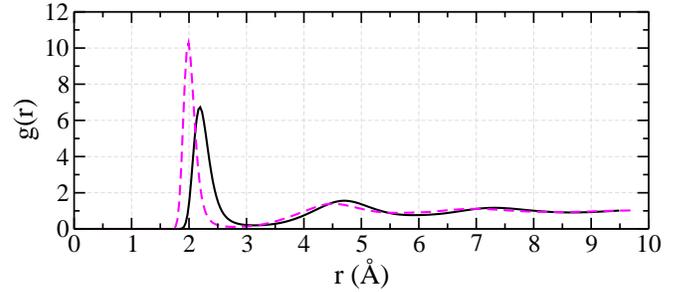}
\caption{\label{fig:rdfZrY}
Radial distribution functions for Y$^{3+}$--F$^-$ (solid line in black) and Zr$^{4+}$--F$^-$ (dashed line in magenta) respectively in LiF-YF$_3$ at 1130~K and $x_{\text{YF}_{3}}=0.15$ and LiF-ZrF$_4$ at 1200~K and $x_{\text{ZrF}_{4}}=0.35$. The RDF for Zr$^{4+}$--F$^-$ is taken from Pauvert et al.~\cite{pauvert2011a}.}
\end{figure}

In Fig.~\ref{fig:rdf}, we present the partial radial distribution
functions (RDF) $g_{ij}\left(r\right)$ for F$^{-}$-F$^{-}$, F$^{-}$-Y$^{3+}$, Y$^{3+}$-Y$^{3+}$ and Li$^+$-F$^{-}$ pairs at 1130~K. Situations corresponding to low and high contents in YF$_{3}$
($x_{\text{YF}_{3}}=0.15$ and 0.60) are presented. Firstly, we observe that the Y-F ionic bond length, taken at the position
of the first maximum in $g(r)$, is of $\approx$2.24~\AA\ for all the compositions. This is significantly larger than the Zr-F distance ($\approx$2.02~\AA) in fluorozirconate melts shown in Fig.~\ref{fig:rdfZrY} while both cations have rather similar short-range repulsion potentials~\cite{salanne2012b}. This is due to the larger charge carried by the Zr$^{4+}$ cations which therefore tend to be more attractive to fluorides with opposite charges. Secondly, we observe that the first peak in this Y-F RDF is 1.5 times less intense
and sharper than its Zr-F counterpart. Also, the following minimum
is less marked in YF$_{3}$. These quantities seem to show that the
first solvation shell of Y$^{3+}$ is more weakly bound than the one of Zr$^{4+}$ in molten fluorides. In the case of the Li$^+$-F$^-$ pair, the RDF has a shape very similar to the case of other molten fluoride mixture~\cite{salanne2006a}.

The extraction of precise coordination numbers
is a difficult experimental task, for example by extended X-ray absorption
fine structure (EXAFS) spectroscopy, that leads to errors of the order of 20~\% for liquid state systems.
 Here the difficulty is increased by the high temperatures of the systems. From NMR, it is possible to extract an average coordination number by comparing the chemical shift of the probed species to typical values obtained for this quantity in well-defined solid state compounds. This was performed in LiF-YF$_3$ mixtures~\cite{rollet2004b,bessada2009a}, which led to the conclusion that the yttrium ions are most probably 7-fold or 8-fold coordinated. A smaller coordination number (CN) of 6 was extracted from Raman spectroscopy experiments by Dracopoulos \text{et al.} in KF-YF$_3$ mixtures~\cite{dracopoulos1997a,dracopoulos1998a}. Such a reduction in the CN when switching from LiF to KF-based systems and a stabilization of octahedral coordination shell in the latter is consistent with our recent results in molten fluorozirconates~\cite{pauvert2011a}.

\begin{figure}
\includegraphics[width=1\columnwidth]{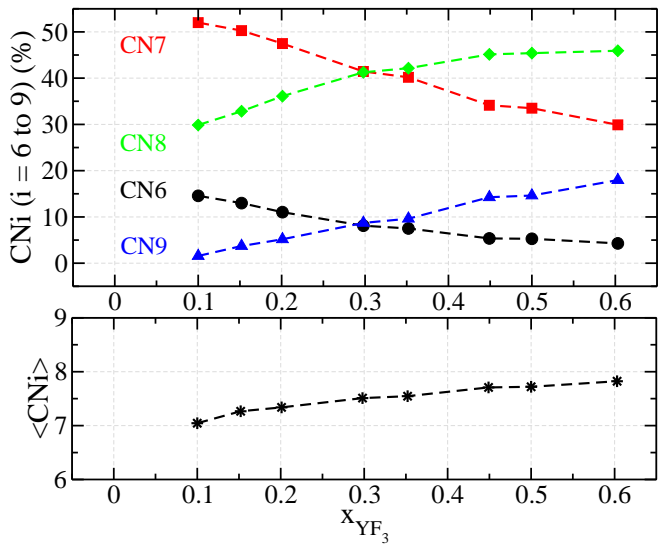}
\caption{
(top) Evolution of each coordination numbers with the composition
of the melt. CN$i$ stands for the proportion of $i$-fold coordinated
ytrium ions. No noticeable differences were observed between the two temperatures studied here (1130~K and 1200~K). Reported values have a typical error bar of 1~\%.\\(Bottom) Average coordination number of ytrium ions as a function of the composition of the melt.\label{fig:coordination}}
\end{figure}

Because of the intrinsic atomic-scale of our MD simulations, we can extract
CNs with estimated error less than 2~\%. 
The evolution of the CN of yttrium with its
molar fraction in LiF-YF$_{3}$ is presented in Fig.~\ref{fig:coordination}.
We find the predominant species to be 7-fold coordinated yttrium ions for $x_{\text{YF}_{3}}$ below 30~\%
and 8-fold coordinated ions above. From NMR experiments, one extracts the average CN over time. From Fig.~\ref{fig:coordination}, the average CN for an yttrium ion at the eutectic composition ($x_{\text{YF}_3} = 0.2$) is $\approx 7.4$ in our simulations, which compares well with the one estimated from NMR experiments~\cite{rollet2004b,bessada2009a}. We observe almost linear evolutions for
all CNs. The proportions of CNs 6 and 7 decrease with $x_{\text{YF}_{3}}$, and are compensated
by increases in CNs~8 and 9. Contrarily to the case of LiF-ZrF$_4$ mixtures, no anomalous behaviour is observed around a given specific composition~\cite{pauvert2010a}. The average coordination number $\left< CN_i \right>$, plotted in the bottom panel of Fig.~\ref{fig:coordination}, varies indeed remarkably slowly and linearly with $x_{\text{YF}_{3}}$, between 7.0 and 7.8.

As for the like-like RDFs which are also shown in Fig.~\ref{fig:rdf}, the F-F one does not show any strong evolution with the composition of the melt. The Y-Y one is characterized by a first peak, which is located at a distance of $\approx$3.9~\AA\, which is smaller than twice the Y-F distance and therefore corresponds to pairs of cations that share a common fluoride anion. The intensity of this first peak slowly increases with the YF$_3$ concentration. This evolution is linked to the slow increase in average CN discussed just above and shown in the bottom panel of Fig.~\ref{fig:coordination}.  It is larger than the one of the corresponding Zr-Zr RDF in fluorozirconates~\cite{pauvert2010a,pauvert2011a,rollet2012b}. Such a structural feature, which denotes some medium-range, network-like ordering, has been shown to impact importantly the transport properties in molten fluorides. 

\section{Discussion: interplay between structure and dynamics}
In molten salts, the coordination number is often seen as the most important structural criterion. It is heavily used for the interpretation of various properties through the perspective of the Lux-Flood acidity~\cite{lux1939a,flood1947a}. In a molten fluoride melt, this definition of acidity induces that (i) an acid is a fluoride ion acceptor and (ii) a base is a fluoride ion donor~\cite{bieber2011a}. In this context, based on their coordination numbers, which range between 6 and 9, Y$^{3+}$ and Zr$^{4+}$ should both be considered as strongly acidic species. Nevertheless, very distinct behaviours are observed in these melts concerning the transport properties.

On the one hand, in MF-ZrF$_4$ melts (with $M^+$ being Li$^+$, Na$^+$, K$^+$ or even a mixture of Li$^+$ and Na$^+$), the $D_{\rm F}/D_{\rm M}$ ratio starts from a value very close to unity in the alkali fluoride melt, then strongly decreases upon addition of ZrF$_4$. In parallel, the $D_{\rm F}/D_{\rm Zr}$ ratio also decreases, taking values close to unity in ZrF$_4$ concentrated melts~\cite{salanne2009b,pauvert2011a} (note that a similar behavior is observed in LiF-BeF$_2$ mixtures~\cite{salanne2006a,salanne2007b}). Here, in LiF-YF$_3$ both ratios hardly vary, showing that the diffusion of fluoride is barely impacted by the presence of yttrium ions, even at high YF$_3$ concentrations.  

On the other hand, we have already mentioned that the Nernst-Einstein approximation overestimates the electrical conductivity of LiF-YF$_3$ molten salts by
10 to 20~\%, independently of the composition. Such a relatively good agreement is usually observed in molten alkali halides, \textit{i.e.}, without complexing species. In LiF-NaF-ZrF$_4$ mixtures, we have shown that deviation from the Nernst-Einstein approximation increases with $x_{\rm ZrF_4}$, the content of complexing species. For the highest concentration studied ($x_{\rm ZrF_4}$=0.29) the conductivity is overestimated by as much as a factor of 2 if correlations between the ions are neglected, in agreement with the picture of Margulis et al. discussed above that relates strong correlations in mobilities of the various ions in molten salts to a decrease in conductivity\cite{kashyap_how_2011}. This again shows that association effects between Y$^{3+}$ and F$^-$ ions are weak in LiF-YF$_3$ melts, despite the high coordination numbers. These points lead to the picture of a loosely bound network structure.
The association behaviour of LiF-YF$_3$ is thus qualitatively different from that of LiF-ZrF$_4$, although their coordination number is high.

In order to investigate further the associative behaviour in LiF-YF$_3$, we analyze the rate of change of the instantaneous coordination number,
which can be interpreted as the solvation shell lifetime. This is done by performing a cage correlation function
analysis. The first minimum of the Y-F partial radial distribution
function is used as a simple geometrical criterion to determine the limit of the first coordination shell of a given Y$^{3+}$. The number of counter-ions F$^-$
that have left the shell between times $t$ and $t+\delta t$
is noted $n_{i}^{out}\left(t,t+\delta t\right)$. The rate at which
a single counter-ion (F$^{-}$) leaves the Y$^{3+}$ ions solvation
shell (the cage) is given by the cage-out correlation function~\cite{rabani1997a}, which is defined as:
\begin{equation}
C^{\text{out}}_\alpha\left(t\right)=\left\langle \Theta\left(1-n_{i}^{\text{out}}\left(0,t\right)\right)\right\rangle,
\end{equation}
where $\Theta$ is the Heaviside step function.

\begin{figure}
\includegraphics[width=1\columnwidth]{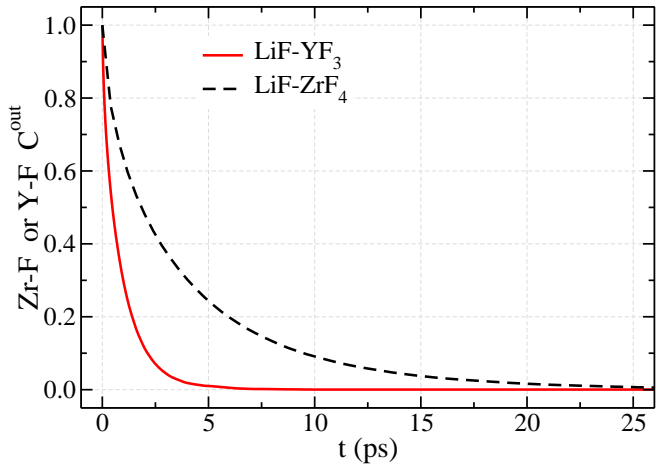}
\caption{\label{fig:CageCF}Cage-out correlation functions for Y$^{3+}$ and
Zr$^{4+}$ ions solvation shell in LiF-YF$_{3}$ and LiF-ZrF$_{4}$,
respectively, at 1200~K and molar fraction $x_{\rm YF_{3}}$ or $x_{\rm ZrF_{4}}=0.35$.}
\end{figure}

The cage-out functions for $\alpha=$Y$^{3+}$ and Zr$^{4+}$ in LiF-YF$_{3}$ and LiF-ZrF$_{4}$
are presented in Fig.~\ref{fig:CageCF} for the same molar fraction of 0.35 at 1200~K. Note that the changes with composition are very small, as shown in the supplementary materials. The exponential decay of the cage-out function is much faster for the Y$^{3+}$-based salt. The corresponding characteristic time is of 0.7~ps, which is much lower than the ones obtained in
LiF-ZrF$_{4}$ (3.1~ps), NaF-ZrF$_{4}$ (15.7~ps) and KF-ZrF$_{4}$ (76.4~ps). These values are summarized in Table~\ref{tab:time_Ea}. The very short characteristic time measured in the case of LiF-YF$_3$ is at the origin of the particular transport properties obtained in this particular molten salt. It shows that even if Y$^{3+}$ and Zr$^{4+}$ have rather similar (and high) coordination numbers, the ionic bond is much weaker for the former.

\begin{figure}
\includegraphics[width=1\columnwidth]{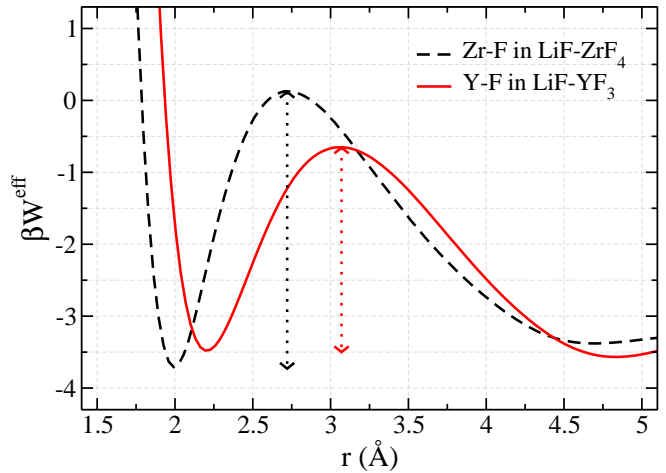}
\caption{\label{fig:pmf}Effective potential between Zr-F and Y-F ionic pairs, at 1200~K and molar fraction $x_{\rm ZrF_{4}}$ or $x_{\rm YF_{3}}=0.35$. Vertical dotted lines indicate the height of the barrier according to the transition state theory.}
\end{figure}

Although the use of such cage correlation functions appears to be very useful for understanding the associative properties of molten salts, this quantity can only be obtained from simulations. It would nevertheless be valuable to extract similar information from a quantity which is more readily accessible. To this end, we compare in Fig.~\ref{fig:pmf} the effective potential between Zr$^{4+}$ and F$^-$, and between Y$^{3+}$ and F$^-$, in the two melts. These were obtained from:
\begin{equation}
\beta W_{ij}^{\rm eff}(r)=-\ln g_{ij}(r)-2 \ln(r)\label{eq:pmf}.
\end{equation}
Indeed, Eq.~\ref{eq:pmf} relates the partial radial distribution functions $g_{ij}(r)$ discussed above to an effective interaction potential. The last term in Eq.~\ref{eq:pmf} accounts for the rotational invariance of $W_{ij}^{\text{eff}}(r)$. Following the transition state theory~\cite{truhlar1983a,masia2003a}, the rate at which the F$^-$ ions escape from  the Y$^{3+}$ solvation shell should depend strongly on the height of the corresponding barrier. We observe a smaller barrier (by $\approx$ 1~$k_B T$) in the LiF-YF$_3$ melt, in agreement with the smaller characteristic time extracted from the cage-out correlation function. This clearly shows that the shape of the RDF, rather than the CN distribution, should be studied for determining the associating properties in a molten salt.

These functions show a deep first minimum followed by a local maximum. The height of this variation can be seen as an activation energy ($E_a$), that we tabulate for LiF-YF$_{3}$, LiF-ZrF$_{4}$, NaF-ZrF$_{4}$ and KF-ZrF$_{4}$ in Table~\ref{tab:time_Ea}. This activation energy ($E_a$) that is the thermal energy a single counterion (F$^-$) has to provide to cage out, is minimal in the case of Y$^{3+}$. This also strenghtens the conclusions about the loosely bound nature of LiF-YF$_3$.
\begin{table}
\begin{tabular}{|c|c|c|}
\hline 
 & $\tau$~(ps) & $E_{a}$~($k_{B}T$)\\
\hline 
\hline 
LiF-YF$_{3}$ & 0.7 & 3.02\\
\hline 
LiF-ZrF$_{4}$ & 3.1 & 4.48\\
\hline 
NaF-ZrF$_{4}$ & 15.7 & 5.39\\
\hline 
KF-ZrF$_{4}$ & 76.4 & 6.72\\
\hline 
\end{tabular}
\caption{\label{tab:time_Ea} Lifetime ($\tau$) of the Y$^{3+}$ first solvation shell extracted
from the cage-out correlation function and activation energies ($E_{a}$)
extracted from the potential of mean forces.}
\end{table}

\section{Conclusion}

We have characterized the transport properties of LiF-YF$_3$ melts by combining PFG-NMR spectroscopy, potentiometric measurements and molecular dynamics simulations. The latter are able to provide diffusion coefficients of Li$^+$ and F$^-$ ions and electrical conductivities in quantitative agreement with the experiments. It is then possible to extract many other quantities, such as the partial radial distribution functions, the coordination number distributions and the cage-out correlation functions, as a function of the molar fraction of YF$_3$. We have shown that despite being appealing, the coordination number is \emph{not} the right observable to predict the strength of the association between a metallic species and the anion. It does therefore not bring much insight on the transport properties of molten salts. Indeed, although Y$^{3+}$ and Zr$^{4+}$ ions have similar coordination numbers in fluorozirconate melts, their diffusion coefficients do not show similar behavior: In LiF-ZrF$_4$ melts, when $x_{\rm ZrF_4}$ increases, $D_{\rm F}$ decays much faster than $D_{\rm Li}$, eventually reaching $D_{\rm Zr}$. On the contrary, in LiF-YF$_3$ melts, the three diffusion coefficients evolve in a similar way.

The analysis of the effective potentials, which can readily be extracted from the radial distribution functions, and of the cage correlation functions, which measure the lifetime of the first solvation shell of the ions, provide a much more complete framework for interpreting the dynamic properties. Here we show that the variation of the diffusion coefficients is linked to a short lifetime of the Y-F ionic bonds, which in turn arises from a rather shallow well in the effective potential. This points toward the potential use of the transition state theory for rationalizing the ionic association effects in molten salts.
 
\begin{acknowledgments}
The authors acknowledge the support of the French Agence Nationale de la Recherche (ANR), under grant ANR-09-BLAN-0188 (``MILIFOX'') and of the PACEN program. The research leading to these results has received funding from the European Community's
Seventh Framework Program under grant agreement n$^{\rm o}$ 249696 EVOL.
\end{acknowledgments}


%

\end{document}